\documentclass[USenglish,twocolumn]{article}

\ifx\directlua\undefined\ifx\XeTeXcharclass\undefined
  \usepackage[utf8]{inputenc}                           
  \else\RequirePackage[no-math]{fontspec}[2017/03/31]\fi 
  \else\RequirePackage[no-math]{fontspec}[2017/03/31]\fi 
\usepackage[sort&compress,square,numbers]{natbib}
\usepackage[big,online]{dgruyter}

\usepackage{color}

\theoremstyle{dgthm}

\theoremstyle{dgdef}

\begin{document}

\articletype{Research Article}

\author*[1]{Iris Niehues}
\author[2]{Daniel Wigger}
\author[3]{Korbinian Kaltenecker} 
\author[4]{Annika Klein-Hitpass}
\author[5]{Philippe Roelli}
\author[6]{Aleksandra K. D\k{a}browska}
\author[7]{Katarzyna Ludwiczak}
\author[8]{Piotr Tatarczak}
\author[9]{Janne O. Becker}
\author[10]{Robert Schmidt}
\author[11]{Martin Schnell}
\author[12]{Johannes Binder}
\author[13]{Andrzej Wysmo\l{}ek}
\author[14]{Rainer Hillenbrand}

\affil[1]{Institute of Physics, University of Münster, Wilhelm-Klemm-Str. 10, 48149 Münster, Germany, iris.niehues@uni-muenster.de; https://orcid.org/0000-0001-7438-2679}
\affil[2]{Department of Physics, University of Münster, Wilhelm-Klemm-Str. 9, 48149 Münster, Germany; https://orcid.org/0000-0002-4190-8803}
\affil[3]{Chair in Hybrid Nanosystems, Nano-Institute Munich, Department of Physics, Ludwig-Maximilians-Universität München, Königinstr. 10, 80539 Munich, Germany, attocube aystems AG, Eglfinger Weg 2, 85540 Haar, Germany; https://orcid.org/0000-0002-9831-6075}
\affil[4]{Institute of Physics, University of Münster, Wilhelm-Klemm-Str. 10, 48149 Münster, Germany}
\affil[5]{CIC nanoGUNE BRTA, Tolosa Hiribidea 76, 20018 Donostia-San Sebastián, Spain; https://orcid.org/0000-0002-1582-2301}
\affil[6]{Faculty of Physics, University of Warsaw, ul. Pasteura 5, 02-093, Warsaw, Poland; https://orcid.org/0000-0003-4633-2054}
\affil[7]{Faculty of Physics, University of Warsaw, ul. Pasteura 5, 02-093, Warsaw, Poland; https://orcid.org/0000-0002-7279-4669}
\affil[8]{Faculty of Physics, University of Warsaw, ul. Pasteura 5, 02-093, Warsaw, Poland; https://orcid.org/0000-0003-3715-9165}
\affil[9]{Institute of Physics, University of Münster, Wilhelm-Klemm-Str. 10, 48149 Münster, Germany}
\affil[10]{Institute of Physics, University of Münster, Wilhelm-Klemm-Str. 10, 48149 Münster, Germany; https://orcid.org/0000-0002-8856-3347}
\affil[11]{CIC nanoGUNE BRTA, Tolosa Hiribidea 76, 20018 Donostia-San Sebastián, Spain, IKERBASQUE, Basque Foundation for Science, 48013 Bilbao, Spain; https://orcid.org/0000-0003-3514-3127}
\affil[12]{Faculty of Physics, University of Warsaw, ul. Pasteura 5, 02-093, Warsaw, Poland; https://orcid.org/0000-0002-0461-7716}
\affil[13]{Faculty of Physics, University of Warsaw, ul. Pasteura 5, 02-093, Warsaw, Polandś https://orcid.org/0000-0002-8302-2189}
\affil[14]{CIC nanoGUNE BRTA, Tolosa Hiribidea 76, 20018 Donostia-San Sebastián, Spain, IKERBASQUE, Basque Foundation for Science, 48013 Bilbao, Spain, Department of Electricity and Electronics, UPV/EHU, 20018 Donostia-San Sebastián, Spainś https://orcid.org/0000-0002-1904-4551}

\title{Nanoscale resolved mapping of the dipole emission of hBN color centers with a scattering-type scanning near-field optical microscope}

\runningtitle{Near-field color center PL}

\abstract{Color centers in hexagonal boron nitride (hBN) are promising candidates as quantum light sources for future technologies. In this work, we utilize a scattering-type near-field optical microscope (s-SNOM) to study the photoluminescence (PL) emission characteristics of such quantum emitters in metalorganic vapor phase epitaxy grown hBN. On the one hand, we demonstrate direct near-field optical excitation and emission through interaction with the nanofocus of the tip resulting in a sub-diffraction limited tip-enhanced PL hotspot. On the other hand, we show that indirect excitation and emission via scattering from the tip significantly increases the recorded PL intensity. This demonstrates that the tip-assisted PL (TAPL) process efficiently guides the generated light to the detector. We apply the TAPL method to map the in-plane dipole orientations of the hBN color centers on the nanoscale. This work promotes the widely available s-SNOM approach to applications in the quantum domain including characterization and optical control.}


\journalname{Accepted version of DOI: 10.1515/nanoph-2024-0554, Nanophotonics}

\journalyear{(2025)}

\maketitle

\section{Introduction}
Color centers in hexagonal boron nitride (hBN) have emerged as important quantum light sources due to their stable and bright single-photon emission at room temperature~\cite{tran2016quantum,tran2016quantum_2,grosso2017tunable,wigger2019phonon} as well as their compatibility with photonic and electronic technologies~\cite{kim2018photonic,kim2019integrated,elshaari2021deterministic,li2021integration,preuss2022low}. Due to these properties, they are promising candidates for applications in quantum communication, quantum computation, and sensing technologies, making the understanding and manipulation of their properties crucial.
Scattering-type near-field optical microscopy (s-SNOM) is an advanced imaging technique that surpasses the diffraction limit, facilitating optical measurements down to the nanoscale~\cite{keilmann2004near,chen2019modern}. Utilizing sharp metallic tips of an atomic force microscope (AFM), s-SNOM not only provides topographical data but also enhances optical contrasts of local material properties~\cite{ocelic2006pseudoheterodyne}. Further, the combination with techniques such as tip-enhanced Raman spectroscopy (TERS)~\cite{bailo2008tip,zhang2013chemical,sonntag2014recent,zhang2016tip,hoppener2024tip} and photoluminescence (TEPL)~\cite{park2016hybrid,yang2020sub,lee2020tip,hasz2022tip,albagami2022tip} extend the capabilities of s-SNOM~\cite{kusch2018combined,fali2021photodegradation,garrity2022probing}, making it ideal for examining the unique photophysical properties of single-photon emitters (SPEs) in materials like hBN. TEPL has so far been used to measure the quantum efficiency~\cite{nikolay2019direct} or for nanoscale imaging with resonant nanoantennas~\cite{palombo2020nanoscale} of hBN emitters. Specially designed tips with resonant plasmonic particles have also been applied to study single molecules~\cite{frey2004high,anger2006enhancement,taminiau2008optical,singh2014vectorial} and quantum dots~\cite{farahani2007bow,gross2018near}.

\section{Results}
In this study, we examine the dipole emission characteristics of color centers embedded in 30 nm thick hBN layers grown by metalorganic vapor phase epitaxy~\cite{chugh2018flow,dabrowska2020two,tokarczyk2023effective,binder2023epitaxial} (MOVPE; see supplementary material (SM) Sec.~S1 for details). The layers are grown on sapphire and transferred onto a gold substrate~\cite{ludwiczak2024large}. Our investigation utilizes a scattering-type near-field optical microscope ({\it neaSCOPE} from {\it neaspec/attocube}) employing a standard metallized AFM tip (\textit{Arrow-NCPt} sourced from \textit{NanoWorld}) illuminated by monochromatic laser light. 
The tip acts as an optical antenna, transforming the incident p-polarized~\cite{aigouy1999polarization} light into a highly focused near field at the tip apex, the so-called nanofocus. The nanofocus interacts with the sample leading to modified scattering from the tip and encoding local sample properties. By recording tip scattered light as function of sample position (note that the sample is scanned), we obtain nanoscale resolved near-field optical images reaching spatial resolutions down to 30 nm, determined by the tip apex~\cite{keilmann2004near,chen2019modern,kusch2018combined,garrity2022probing}. 
When detecting elastically scattered light, to study the local dielectric function of the investigated medium, the s-SNOM signals are demodulated with respect to the AFM tapping mode frequency~\cite{ocelic2006pseudoheterodyne}. However, when studying inelastically scattered light, which in our case is the PL from hBN color centers, measurements are usually performed without signal demodulation due to the low count rates, but demodulation is possible under special conditions~\cite{gerton2004tip}.
Note, that our s-SNOM setup includes a high-quality, silver-protected off-axis
parabolic mirror with a numerical aperture (NA) of 0.72, which optimizes the focusing and collection efficiency of the optical system and is crucial for the performed PL measurements. 

\subsection{Characterization of photoluminescence mapping}\label{sec:2_1}
In our specific experiments, we employ the near-field optical microscope in tapping mode, with low oscillation amplitudes between 20 nm and 30 nm, to detect PL signals influenced by the presence of the metallic AFM tip (a detailed description of the measurement procedures is given in the SM Sec.~S2). Note that we use standard metallic AFM tips (\textit{Arrow-NCPt} sourced from \textit{NanoWorld}).
Throughout this study, we use a 532~nm (2.33~eV) laser for the optical excitation of the hBN color centers (examples with other wavelengths are given in the SM Sec.~S3, Fig.~S1).

\begin{figure*}[ht!]
\centering\includegraphics[width=0.75\textwidth]{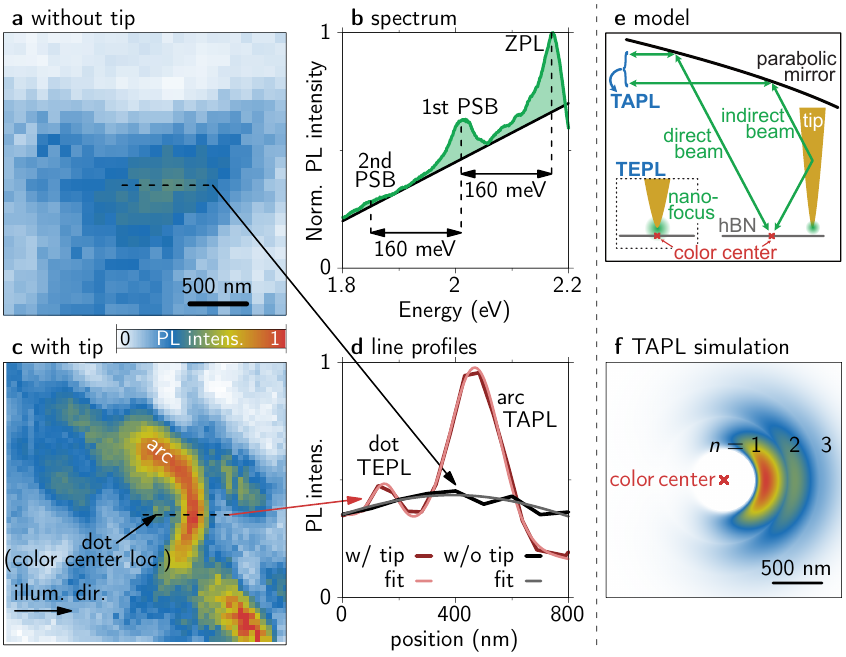}
\caption{PL measurement of a single color center taken with an AFM tip. The images are shown with the same color bar for better comparison of the observed PL intensities. (a) PL intensity map without the tip showing a diffraction limited emission spot. (b) PL spectrum of the studied emitter recorded with an extended integration time inside the arc in (c). The zero-phonon line (ZPL) and optical phonon sidebands (PSBs) of 160 meV are marked as well as the broad background PL (black line). (c) PL map of the same emitter with the AFM tip showing two sub-diffraction limit features marked as ‘dot’ and ‘arc’. (d) Line profiles along the dashed lines in (a) in black and (c) in red (dark measurement, bright Gaussian fits). The fitted full widths at half maximum (FWHM) are 110~nm (dot), 209~nm (arc) and 1418~nm (w/o tip). (e) Schematic of the interference between direct and indirect excitation/emission of the color center via the AFM tip (TAPL). Inset shows the nanofocus interaction at the location of the color center explaining the dot (TEPL). (f) Analytical reproduction of the TAPL arc in (c) applying the model in (e).}\label{fig:1}
\end{figure*}

Figure~\ref{fig:1}a presents a typical far-field PL image of a single hBN color center recorded in the near-field optical microscope without a tip, displaying an elliptical maximum in the PL map with extensions of roughly 1 \textmu m $\times$ 4 \textmu m. This  gives a benchmark for the optical resolution of the setup without AFM tip. To confirm that the light emission stems from a single color center, the corresponding PL spectrum is depicted in Fig.~\ref{fig:1}b, which shows the typical asymmetric zero-phonon line (ZPL) and phonon sideband (PSB) characteristics~\cite{wigger2019phonon,preuss2022resonant}. Note that we did not analyze the statistical properties of the light emission, since for our present study it is not of relevance whether single-photon or classical emission is measured. To record PL maps, at each pixel a PL spectrum is taken and we integrate the intensity of the green shaded area in Fig.~\ref{fig:1}b. Note that we determine this area under the peaks in practice by fitting as described in Ref.~[4]. This procedure is performed for all PL maps shown in this work. The black line in Fig.~\ref{fig:1}b indicates the broad PL background homogeneously observed over the whole MOVPE grown sample~\cite{dabrowska2024defects}.

Notably, the PL map of the color center from Fig.~\ref{fig:1}a experiences a dramatic transformation when imaged with a metallic AFM tip, as shown in Fig.~\ref{fig:1}c (see also the direct overlay of the two images in SM Fig.~S2). Here, the laser is focused on the tip generating the near-field nanofocus on the size of the tip apex (typically around 30~nm). Consequently, the sample is simultaneously excited via the far-field focus and the localized near-field focus. The tip is scanned over the sample and PL spectra are recorded at each position (pixel). 
The PL map in Fig.~\ref{fig:1}c reveals two distinct features: (I) a circular symmetric emission hotspot which we will refer to as ‘dot’ and (II) a more pronounced ‘arc’ around the dot. It is important to note that this specific dot+arc pattern is consistently observed across many imaged hBN color centers (see detailed discussion in SM Sec.~S3).
We will discuss the origins of the two features (I: dot and II: arc) in the following paragraphs.
 
(I) Origin of the dot structure: 
To quantify the spatial resolution when recording PL maps with the AFM tip, Fig.~\ref{fig:1}d shows a line profile along the dashed line in (c) (dark red) and a fit with two Gaussians on a tilted background (bright red). From this we extract a full width at half maximum (FWHM) of 110~nm for the dot, which is clearly below the diffraction limit for far-field experiments. Therefore, we identify the dot as a result of the direct near-field interaction between the nanofocus at the tip apex and the color center leading to TEPL. The inset of Fig.~\ref{fig:1}e schematically illustrates this explanation. For reference, the black line in Fig.~\ref{fig:1}d shows a line profile along the dashed line in Fig.~\ref{fig:1}a (without AFM tip) and the Gaussian fit reveals a FWHM of around 1400~nm being the resolution in the setup without tip.
 
(II) Origin of the arc structure: The prominent arc structure has a diameter of around 1000~nm and a FWHM cross section of 209~nm (extracted from the line profile and fit in Fig.~\ref{fig:1}d). We can explain this feature by constructive interference between direct beams to/from the color center and those scattered from the AFM tip (indirect beam) as sketched in Fig.~\ref{fig:1}e. To distinguish this effect from the near-field interaction leading to TEPL (dot), we call it tip-assisted photoluminescence (TAPL). To verify our explanation, we use a simple model to calculate the interference depending on the tip location. Besides the interference condition, the model includes an incidence/collection angle and interference widths, described by Gaussian intensity distributions, around this angle. We further take the NA of the parabolic mirror for the focussing angle and the movement of the color center with respect to the broadened focal point into account. A detailed description of the analytic interference model is given in SM Sec.~S5 and the result for the TAPL signal, i.e., the arc, is shown in Fig.~\ref{fig:1}f. We find an overall good agreement with the measurement and the higher interference orders leading to the arc replicas ($n=2,3$) show up in measurements with high optical powers (see Fig.~\ref{fig:2}). This handy description gives an alternative explanation to the formation of standing surface waves between tip and emitter, suggested in Ref.~\cite{jo2023direct}. In our model the effect does not depend on the metallic surface of the substrate. Indeed, we find similar PL maps of emitters in hBN transferred onto Si/SiO$_2$ substrates (see SM Fig.~S1c). Note, that the direct nanofocus interaction with the color center resulting in the dot, i.e. TEPL, is not included in the model.

To finish this initial characterization of our PL maps, we come back to the experiment in Fig.~\ref{fig:1}c and have a closer look at the intensities of the TEPL and TAPL features, dot and arc, respectively. 
Here and in all measurements we find that the TEPL intensity of the dot is significantly weaker than the TAPL intensity of the arc. 
Thinking of the term 'enhanced' in TEPL this low intensity in the dot might appear counter-intuitive. On the one hand this could be caused through quenching effects by the proximity of the metallic AFM tip~\cite{anger2006enhancement,palombo2020nanoscale}. On the other hand the weak intensity of the dot could stem from averaging of the PL signal during the vertical tip movement in tapping mode. The tip oscillates with an amplitude of 20~nm and we average the PL signal for 0.5~s. Therefore, the TEPL signal only contributes when the tip is in close proximity to the color center, and rapidly diminishes for higher tip positions owing to the exponential distance dependence of the tip's near field. This results in a reduced time-averaged TEPL signal. 
Conversely, the interference between the direct and indirect beams in TAPL (arc) depends only weakly on the tip height, thus contributing strongly to the overall PL signal during the whole tip oscillation cycle, resulting in the rather strong TAPL signal of the arc.

Finally, we compare the recorded PL intensities with and without the AFM tip. By spatially summing over the collected light intensity in the PL map in Fig.~\ref{fig:1}a without tip and the PL map in Fig.~\ref{fig:1}c  with the tip (see SM Sec.~S4 for details) we can estimate that we achieve an overall sixfold increased efficiency in this example. Note that the peak intensities differ by a factor of two (see Fig.~\ref{fig:1}d). 
This demonstrates that the interference effect in TAPL due to the presence of the AFM tip contributes to the enhancement via two effects: (A) The optical excitation of the color center becomes more efficient and (B) the PL from the color center is guided efficiently into the collection angle of the parabolic mirror.
Consequently, through this TAPL mapping we can identify sample locations where metal antennas can be positioned to help guiding light towards an absorption center and guiding emitted light efficiently to a detector.

\subsection{Demonstration of TEPL: Bleaching of color centers}
\begin{figure}[ht!]
\centering\includegraphics[width=\columnwidth]{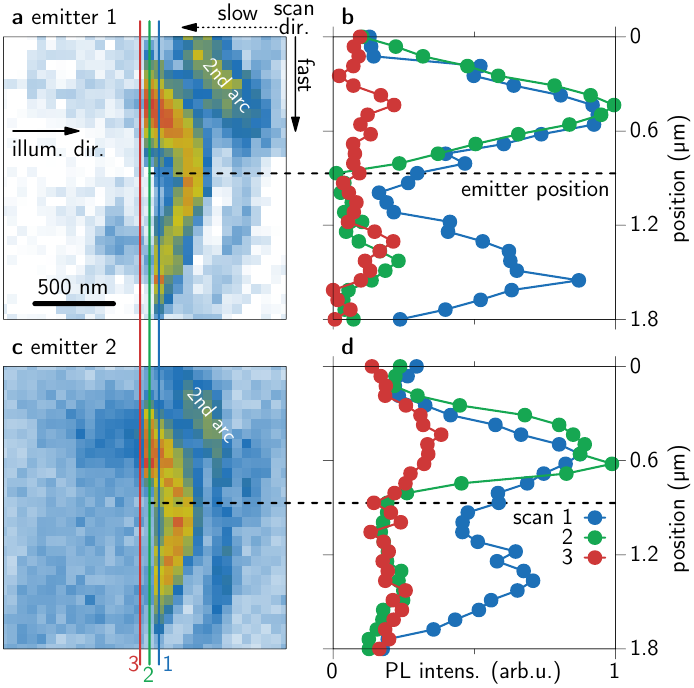}
\caption{Near-field scan of two different emitters (a, c), with line profiles along the marked positions in (b, d). The measurement starts in the top right corner and line scan run vertically ending in the bottom left corner. Along the green scan line, the PL bleaches at the tip location of the emitter (dashed vertical line).}\label{fig:2}
\end{figure}

It is known since the first studies on hBN color centers, that optical excitation with too high intensities results in bleaching of the emission~\cite{li2023prolonged}. In Fig.~\ref{fig:2} we use this fact to demonstrate efficient optical driving through the nanofocus at the tip apex, i.e., TEPL. We show PL maps taken with an AFM tip of two emitters (a, c) with an increased optical excitation power of 500~\textmu W (150~\textmu W is used for the other measurements) and selected line profiles along the scan direction of the AFM in (b, d) marked by the colored lines in (a, c). The spectra are acquired in vertical lines starting from the top right to the bottom left. For both color centers the emission bleaches almost entirely at exactly the position where we expect their locations (dashed black line, position of the dot). The line scan before reaching the emitter (blue) covers the double peak structure of the arc and the following scan after bleaching the emitter (red) is nearly entirely flat. The green scan line across the emitter location marked by the left end of the black dashed line still covers the first maximum of the arc but does not show the second one. This indicates that the emission gets bleached when the near field of the tip directly interacts with the color center. From the dot structure in the PL map of Fig.~\ref{fig:1}c, we can conclude that the emitter interacts with the nanofocus of the tip and that the enhanced field at the tip apex leads to TEPL. By increasing the laser power, both the far- and near-field illumination of the emitter increase, eventually reaching its bleaching threshold. Owing to the field enhancement at the tip apex, the threshold is first surpassed when the emitter comes into the near-field nanofocus below the tip apex. Since bleaching is an irreversible process, unlike quenching, TAPL is observed only before (but not after) the emitter is exposed to the nanofocus.
Note, that in the TAPL situation, i.e., in absence of tip-emitter near-field interaction, the local intensity acting on the emitter is still below the bleaching threshold of the emitter and PL is detected.

\subsection{Utilization of TAPL: In-plane dipole emission mapping}
\begin{figure}[ht!]
\centering\includegraphics[width=\columnwidth]{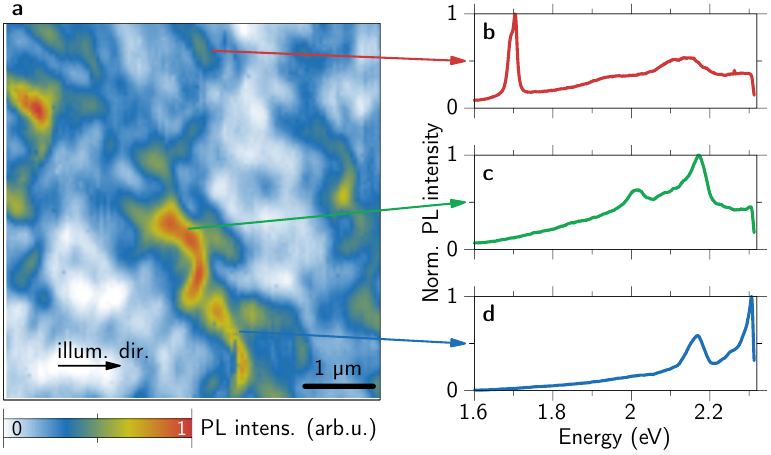}
\caption{(a) Larger PL map with several emitters of different intensities and slightly different arcs, the black arrow marks the illumination direction. (b-d) PL spectra of color centers at the marked positions in the sample with long integration times of 5~s to improve the signal-to-noise-ratio. The spectrum in (c) is the same as in Fig.~\ref{fig:1}b.}\label{fig:3}
\end{figure}

Figure~\ref{fig:3} displays a PL overview map taken with an AFM tip spanning 6~\textmu m $\times$ 6 \textmu m with a pixel size of 50~nm. This map captures several emission centers, each distinguished by the characteristic arc. The same map with marked color center locations according to the arcs is given in SM Fig.~S5. Note that the dot structure associated with TEPL is not visible for every color center. One reason could be that the emitter is located deep in the sample and cannot be reached with the near-field nanofocus.
The PL spectra of three exemplary color centers recorded with long 5~s integration times are plotted in Fig.~\ref{fig:3}b-c. The most prominently visible color center is located in the middle, which is the same as studied in Fig.~\ref{fig:1}. A corresponding AFM height profile is available in SM Fig.~S6. From this measurement we cannot identify a correlation between the morphology of the sample and the appearance of color centers. Note, that from the variation in TAPL intensities we cannot conclude different overall brightnesses of the color centers. As we will show in the following, the main reason is their variation in dipole orientation with respect to the illumination direction.

In the following, we demonstrate how the TAPL maximum can be utilized to determine the in-plane dipole orientation of a single color center in hBN. By systematically rotating the sample relative to the illumination direction we monitor changes in the same emitters’ brightness. The respective AFM images from which we determine each rotation angle are shown in SM Fig.~S7. The color centers are identified in each measurement by their location in the AFM and PL images and their spectral shape/ZPL energy. For each sample rotation, we record a PL map with the AFM tip and analyzed the position and spectra of the arc from four different color centers. For better comparability we use the same tip for the entire measurement series.

\begin{figure}[h!]
\centering\includegraphics[width=\columnwidth]{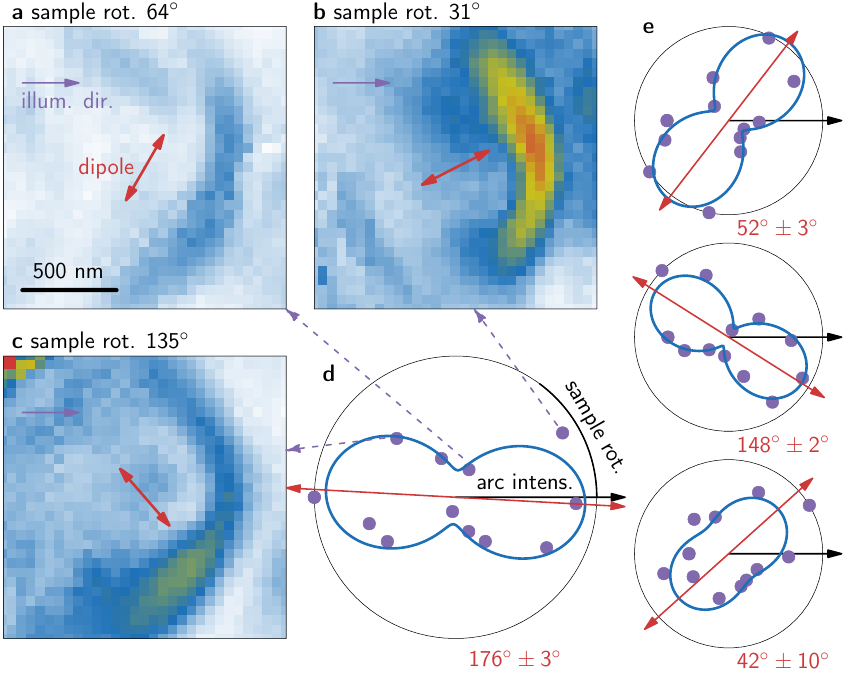}
\caption{Mapping of the dipole orientation by sample rotation. (a-c) PL intensity maps (integrated peak area from fit) for different angles between illumination direction (violet arrows) and sample orientation (red arrows). (d) Polar plot of maximum PL intensities in the arc  as a function of sample rotation (violet dots) with dipole fit (blue curve). (e) Dipole emission patterns of three other color centers in the same sample.}\label{fig:4}
\end{figure}

In Figs.~\ref{fig:4}a-c we show three example PL maps of the emitter from Fig.~\ref{fig:1}c with different sample orientations, i.e., different illumination directions. The sample rotation for each map is stated above and the red arrows indicate the dipole orientation (determined in Fig.~\ref{fig:4}d) in relation to the illumination (violet arrows).

We find that the overall integrated emitter’s TAPL intensity of the arc strongly depends on the illumination direction (violet arrows), where (b) shows a large, (c) a medium, and (a) a small intensity. Figure~\ref{fig:4}d displays the extracted peak intensities inside the arc as a function of illumination direction in a polar plot, where the violet dots (experimental data) are fitted with a dipole pattern (blue). The plot reveals a clear dipolar emission characteristic for the color center with an orientation of $176^\circ\pm3^\circ$. The other studied color centers in Fig.~\ref{fig:4}e show different in-plane dipole orientations with $52^\circ\pm3^\circ$, $148^\circ\pm2^\circ$, and  $42^\circ\pm10^\circ$. We added these dipole orientations to SM Fig.~S5. The non-vanishing intensity at orthogonal orientation between illumination direction and color center dipole can be explained by the emitter's out-of plane dipole component known from literature~\cite{noh2018stark,yu2022electrical,zhigulin2023stark}. These results already demonstrate that this method provides a reliable way to map hBN color centers on the nanoscale and simultaneously gain insight in their optical properties. However, to find quantitative correlations between lattice orientation and emitter dipole, a larger set of emitters needs to be investigated in single crystal samples, which goes beyond the scope of this work. While we clearly recognize the arc in all three examples in Figs.~\ref{fig:4}a-c, also the intensity distribution within the arc varies with the illumination angle. This could be related to the dipole orientation with respect to the crystal lattice, but again needs a dedicated systematic study on single crystal samples. In addition, the TAPL signal could be altered by details of the tip geometry (see Figs.~\ref{fig:1}e, f).
From the perspective of positioning nanoantennas near emitters to improve their excitation and PL (see end of Sec.~\ref{sec:2_1}), our TAPL dipole mapping results demonstrate that the antenna needs to be positioned aligned with the in-plane dipole orientation of the source for maximum efficiency.

\section{Conclusions}
In this study, we have demonstrated the utility of a scattering-type near-field optical microscope (s-SNOM), operated in tapping mode, for measuring the photoluminescence (PL) emission of color centers in hBN. The presence of the standard metallic AFM tip has two effects on the PL maps of the light emission centers. (I)~The direct interaction between the near-field nanofocus at the tip apex and the color center leads to a sub-diffraction limited emission spot and tip-enhanced PL (TEPL) at the location of the emitter. This effect may be exploited further in the future when combining PL detection with demodulation techniques that can isolate the pure TEPL signal in this s-SNOM setup~\cite{mangum2008resolving,hoppener2009background}. This development might improve the spatial resolution of TEPL further down to the nanometer scale using standard metallic AFM tips, as done here. Note, that this is currently not feasible in our setup due to the low count rates, calling for further improvement of the used detector, e.g., single-photon counter/photomultiplier tube.
(II) The far-field interference between direct excitation/PL emission of the color center and the beam that is scattered from the metal tip, resulted in a significant increase in detected PL intensity, which we call tip-assisted PL (TAPL). This effect manifested in an arc PL intensity around the color center and demonstrates that the emission direction from the color center can be controlled by the presence of the metal tip. Perspectively, this finding can be used to guide single photons generated by nanostructures via a metallic antenna. In addition, we have used the TAPL signal to map the dipole orientation of hBN color centers. While this is also possible with far-field methods, our near-field approach works on the scale of a few hundred nm, below the considered wavelength (see Fig.~\ref{fig:1}d). 

We note that the commercial s-SNOM setup used in this work is not uniquely optimized for TEPL measurements~\cite{frey2004high,taminiau2008optical,palombo2020nanoscale}, but was originally designed for IR s-SNOM and nano-FTIR measurements~\cite{hillenbrand2002phonon,keilmann2004near,ocelic2006pseudoheterodyne,huth2012nano}. Our implementation of TEPL and TAPL (possible due to the high NA parabolic mirror), and the demonstration of novel functionality in the form of dipole mapping, renders valuable progress in the development of a s-SNOM operating with multi-spectral and multi-messenger nanoprobes~\cite{mcleod2020multi}.

While we used polycrystalline hBN grown by metalorganic vapor-phase epitaxy (MOVPE), which supports a high density of color centers, our technique can also be applied to exfoliated hBN. This would allow to correlate the dipolar emission direction with the crystallographic axes of the material, potentially addressing the question of whether the SPEs originate from defects~\cite{tawfik2017first,kirchhoff2022electronic} or are molecule-related~\cite{neumann2023organic}. In the latter case, no correlation between the lattice orientation and the emission direction would be expected, while real lattice defects should have a strong correlation.

Our introduced methodology opens new avenues for understanding the underlying physics of color centers in hBN and enhances the capabilities of nanoscale optical microscopy. Perspective applications obviously include sub-diffraction limit control of other quantum excitations in semiconductors, e.g., excitons.

\begin{acknowledgement}
We thank Rudolf Bratschitsch and Steffen Michaelis de Vasconcellos for providing access to their confocal PL setup at the University of M\"unster.
\end{acknowledgement}

\begin{funding}
Ministerium f\"ur Innovation, Wissenschaft und Forschung des Landes Nordrhein-Westfalen: NRW-R\"uckkehrprogramm, Deutsche Forschungsgemeinschaft (DFG) Grant no. 467576442. 
Grant CEX2020-001038-M funded by MICIU/AEI /10.13039/501100011033 and Grant PID2021-123949GB-I00 (NANOSPEC) funded by MICIU/AEI /10.13039/501100011033 and by ERDF/EU.
Project PID2020-115221GA-C44 funded by MICIU/AEI/ 10.13039/501100011033.
National Science Centre, Poland grants 2020/39/D/ST7/02811 and 2022/45/N/ST7/03355.
\end{funding}

\begin{authorcontributions}
 I.N. and A.K-H. conducted the near-field experiments.
K.K., I.N., and P.R. optimized the experimental setup.
D.W. performed the simulations.
I.N., J.O.B., and R.S. performed far-field PL measurements.
A.K.D., K.L., P.T., J.B. and A.W. grew the hBN crystals and prepared the samples.
M.S. contributed to the model development.
I.N. and D.W. analyzed the data and prepared the figures.
I.N., D.W. and R.H. wrote the initial manuscript with input from all coauthors.
All authors discussed the results and approved the final version of the manuscript. All authors have accepted responsibility for the entire content of this manuscript and approved its submission.
\end{authorcontributions}

\begin{conflictofinterest}
R.H. is a co-founder of neaspec GmbH, now part of attocube systems AG, a company producing scattering-type scanning near-field optical microscope systems, such as the one described in this article. The remaining authors declare no competing interests.
\end{conflictofinterest}

\begin{informedconsent}
Informed consent was obtained from all individuals included in this study.
\end{informedconsent}

\begin{ethicalapproval}
\end{ethicalapproval}

\begin{dataavailabilitystatement}
The datasets generated during and/or analyzed during the current study are available from the corresponding author on reasonable request.
\end{dataavailabilitystatement}


\begin{thebibliography}{10}
\providecommand{\url}[1]{#1}
\csname url@samestyle\endcsname
\providecommand{\newblock}{\relax}
\providecommand{\bibinfo}[2]{#2}
\providecommand{\BIBentrySTDinterwordspacing}{\spaceskip=0pt\relax}
\providecommand{\BIBentryALTinterwordstretchfactor}{4}
\providecommand{\BIBentryALTinterwordspacing}{\spaceskip=\fontdimen2\font plus
\BIBentryALTinterwordstretchfactor\fontdimen3\font minus
  \fontdimen4\font\relax}
\providecommand{\BIBforeignlanguage}[2]{{%
\expandafter\ifx\csname l@#1\endcsname\relax
\typeout{** WARNING: IEEEtran.bst: No hyphenation pattern has been}%
\typeout{** loaded for the language `#1'. Using the pattern for}%
\typeout{** the default language instead.}%
\else
\language=\csname l@#1\endcsname
\fi
#2}}
\providecommand{\BIBdecl}{\relax}
\BIBdecl

\bibitem{tran2016quantum}
T.~T. Tran, K.~Bray, M.~J. Ford, M.~Toth, and I.~Aharonovich, ``Quantum
  emission from hexagonal boron nitride monolayers,'' \emph{Nat. Nanotechnol.},
  vol.~11, pp. 37--41, 2016.

\bibitem{tran2016quantum_2}
T.~T. Tran, C.~Zachreson, A.~M. Berhane, K.~Bray, R.~G. Sandstrom, L.~H. Li,
  T.~Taniguchi, K.~Watanabe, I.~Aharonovich, and M.~Toth, ``Quantum emission
  from defects in single-crystalline hexagonal boron nitride,'' \emph{Phys.
  Rev. Appl.}, vol.~5, no.~3, p. 034005, 2016.

\bibitem{grosso2017tunable}
G.~Grosso, H.~Moon, B.~Lienhard, S.~Ali, D.~K. Efetov, M.~M. Furchi,
  P.~Jarillo-Herrero, M.~J. Ford, I.~Aharonovich, and D.~Englund, ``Tunable and
  high-purity room temperature single-photon emission from atomic defects in
  hexagonal boron nitride,'' \emph{Nat. Commun.}, vol.~8, p. 705, 2017.

\bibitem{wigger2019phonon}
D.~Wigger, R.~Schmidt, O.~Del Pozo-Zamudio, J.~A. Preu{\ss}, P.~Tonndorf,
  R.~Schneider, P.~Steeger, J.~Kern, Y.~Khodaei, J.~Sperling, S.~Michaelis~de
  Vasconcellos, R.~Bratschitsch, and T.~Kuhn, ``Phonon-assisted emission and
  absorption of individual color centers in hexagonal boron nitride,'' \emph{2D
  Mater.}, vol.~6, no.~3, p. 035006, 2019.

\bibitem{kim2018photonic}
S.~Kim, J.~E. Fr{\"o}ch, J.~Christian, M.~Straw, J.~Bishop, D.~Totonjian,
  K.~Watanabe, T.~Taniguchi, M.~Toth, and I.~Aharonovich, ``Photonic crystal
  cavities from hexagonal boron nitride,'' \emph{Nat. Commun.}, vol.~9, p.
  2623, 2018.

\bibitem{kim2019integrated}
S.~Kim, N.~M.~H. Duong, M.~Nguyen, T.-J. Lu, M.~Kianinia, N.~Mendelson,
  A.~Solntsev, C.~Bradac, D.~R. Englund, and I.~Aharonovich, ``Integrated on
  chip platform with quantum emitters in layered materials,'' \emph{Adv. Opt.
  Mater.}, vol.~7, no.~23, p. 1901132, 2019.

\bibitem{elshaari2021deterministic}
A.~W. Elshaari, A.~Skalli, S.~Gyger, M.~Nurizzo, L.~Schweickert,
  I.~Esmaeil~Zadeh, M.~Svedendahl, S.~Steinhauer, and V.~Zwiller,
  ``{Deterministic integration of hBN emitter in silicon nitride photonic
  waveguide},'' \emph{Adv. Quantum Technol.}, vol.~4, no.~6, p. 2100032, 2021.

\bibitem{li2021integration}
C.~Li, J.~E. Froch, M.~Nonahal, T.~N. Tran, M.~Toth, S.~Kim, and
  I.~Aharonovich, ``{Integration of hBN quantum emitters in monolithically
  fabricated waveguides},'' \emph{ACS Photonics}, vol.~8, no.~10, pp.
  2966--2972, 2021.

\bibitem{preuss2022low}
J.~A. Preu{\ss}, H.~Gehring, R.~Schmidt, L.~Jin, D.~Wendland, J.~Kern, W.~H.~P.
  Pernice, S.~Michaelis~de Vasconcellos, and R.~Bratschitsch, ``{Low-divergence
  hBN single-photon source with a 3D-printed low-fluorescence elliptical
  polymer microlens},'' \emph{Nano Lett.}, vol.~23, no.~2, pp. 407--413, 2022.

\bibitem{keilmann2004near}
F.~Keilmann and R.~Hillenbrand, ``Near-field microscopy by elastic light
  scattering from a tip,'' \emph{Philos. Trans. R. Soc. A}, vol. 362, no. 1817,
  pp. 787--805, 2004.

\bibitem{chen2019modern}
X.~Chen, D.~Hu, R.~Mescall, G.~You, D.~Basov, Q.~Dai, and M.~Liu, ``Modern
  scattering-type scanning near-field optical microscopy for advanced material
  research,'' \emph{Adv. Mater.}, vol.~31, no.~24, p. 1804774, 2019.

\bibitem{ocelic2006pseudoheterodyne}
N.~Ocelic, A.~Huber, and R.~Hillenbrand, ``Pseudoheterodyne detection for
  background-free near-field spectroscopy,'' \emph{Appl. Phys. Lett.}, vol.~89,
  no.~10, 2006.

\bibitem{bailo2008tip}
E.~Bailo and V.~Deckert, ``{Tip-enhanced Raman scattering},'' \emph{Chem. Soc.
  Rev.}, vol.~37, no.~5, pp. 921--930, 2008.

\bibitem{zhang2013chemical}
R.~Zhang, Y.~Zhang, Z.~C. Dong, S.~Jiang, C.~Zhang, L.~G. Chen, L.~Zhang,
  Y.~Liao, J.~Aizpurua, Y.~L. Luo, and J.~G. Hou, ``{Chemical mapping of a
  single molecule by plasmon-enhanced Raman scattering},'' \emph{Nature}, vol.
  498, no. 7452, pp. 82--86, 2013.

\bibitem{sonntag2014recent}
M.~D. Sonntag, E.~A. Pozzi, N.~Jiang, M.~C. Hersam, and R.~P. Van~Duyne,
  ``{Recent advances in tip-enhanced Raman spectroscopy},'' \emph{J. Phys.
  Chem. Lett.}, vol.~5, no.~18, pp. 3125--3130, 2014.

\bibitem{zhang2016tip}
Z.~Zhang, S.~Sheng, R.~Wang, and M.~Sun, ``Tip-enhanced raman spectroscopy,''
  pp. 9328--9346, 2016.

\bibitem{hoppener2024tip}
C.~H{\"o}ppener, J.~Aizpurua, H.~Chen, S.~Gr{\"a}fe, A.~Jorio, S.~Kupfer,
  Z.~Zhang, and V.~Deckert, ``{Tip-enhanced Raman scattering},'' \emph{Nat.
  Rev. Methods Primers}, vol.~4, p.~47, 2024.

\bibitem{park2016hybrid}
K.-D. Park, O.~Khatib, V.~Kravtsov, G.~Clark, X.~Xu, and M.~B. Raschke,
  ``{Hybrid tip-enhanced nanospectroscopy and nanoimaging of monolayer WSe$_2$
  with local strain control},'' \emph{Nano Lett.}, vol.~16, no.~4, pp.
  2621--2627, 2016.

\bibitem{yang2020sub}
B.~Yang, G.~Chen, A.~Ghafoor, Y.~Zhang, Y.~Zhang, Y.~Zhang, Y.~Luo, J.~Yang,
  V.~Sandoghdar, J.~Aizpurua, Z.~Dong, and J.~G. Hou, ``Sub-nanometre
  resolution in single-molecule photoluminescence imaging,'' \emph{Nat.
  Photonics}, vol.~14, no.~11, pp. 693--699, 2020.

\bibitem{lee2020tip}
H.~Lee, D.~Y. Lee, M.~G. Kang, Y.~Koo, T.~Kim, and K.-D. Park, ``Tip-enhanced
  photoluminescence nano-spectroscopy and nano-imaging,'' \emph{Nanophotonics},
  vol.~9, no.~10, pp. 3089--3110, 2020.

\bibitem{hasz2022tip}
K.~Hasz, Z.~Hu, K.-D. Park, and M.~B. Raschke, ``{Tip-enhanced dark exciton
  nanoimaging and local strain control in monolayer WSe$_2$},'' \emph{Nano
  Lett.}, vol.~23, no.~1, pp. 198--204, 2022.

\bibitem{albagami2022tip}
A.~Albagami, S.~Ambardar, H.~Hrim, P.~K. Sahoo, Y.~Emirov, H.~R. Guti{\'e}rrez,
  and D.~V. Voronine, ``Tip-enhanced photoluminescence of freestanding lateral
  heterobubbles,'' \emph{ACS Appl. Mater. Interfaces}, vol.~14, no.~8, pp.
  11\,006--11\,015, 2022.

\bibitem{kusch2018combined}
P.~Kusch, N.~Morquillas~Azpiazu, N.~S. Mueller, S.~Mastel, J.~I. Pascual, and
  R.~Hillenbrand, ``{Combined tip-enhanced Raman spectroscopy and
  scattering-type scanning near-field optical microscopy},'' \emph{J. Phys.
  Chem. C}, vol. 122, no.~28, pp. 16\,274--16\,280, 2018.

\bibitem{fali2021photodegradation}
A.~Fali, T.~Zhang, J.~P. Terry, E.~Kahn, K.~Fujisawa, B.~Kabius, S.~Koirala,
  Y.~Ghafouri, D.~Zhou, W.~Song, L.~Yang, M.~Terrones, and Y.~Abate,
  ``{Photodegradation protection in 2D in-plane heterostructures revealed by
  hyperspectral nanoimaging: The role of nanointerface 2D alloys},'' \emph{ACS
  Nano}, vol.~15, no.~2, pp. 2447--2457, 2021.

\bibitem{garrity2022probing}
O.~Garrity, A.~Rodriguez, N.~S. Mueller, O.~Frank, and P.~Kusch, ``{Probing the
  local dielectric function of WS$_2$ on an Au substrate by near field optical
  microscopy operating in the visible spectral range},'' \emph{Appl. Surf.
  Sci.}, vol. 574, p. 151672, 2022.

\bibitem{nikolay2019direct}
N.~Nikolay, N.~Mendelson, E.~{\"O}zelci, B.~Sontheimer, F.~B{\"o}hm, G.~Kewes,
  M.~Toth, I.~Aharonovich, and O.~Benson, ``Direct measurement of quantum
  efficiency of single-photon emitters in hexagonal boron nitride,''
  \emph{Optica}, vol.~6, no.~8, pp. 1084--1088, 2019.

\bibitem{palombo2020nanoscale}
N.~Palombo~Blascetta, M.~Liebel, X.~Lu, T.~Taniguchi, K.~Watanabe, D.~K.
  Efetov, and N.~F. Van~Hulst, ``Nanoscale imaging and control of hexagonal
  boron nitride single photon emitters by a resonant nanoantenna,'' \emph{Nano
  Lett.}, vol.~20, no.~3, pp. 1992--1999, 2020.

\bibitem{frey2004high}
H.~G. Frey, S.~Witt, K.~Felderer, and R.~Guckenberger, ``High-resolution
  imaging of single fluorescent molecules with the optical near-field of a
  metal tip,'' \emph{Phys. Rev. Lett.}, vol.~93, no.~20, p. 200801, 2004.

\bibitem{anger2006enhancement}
P.~Anger, P.~Bharadwaj, and L.~Novotny, ``Enhancement and quenching of
  single-molecule fluorescence,'' \emph{Phys. Rev. Lett.}, vol.~96, no.~11, p.
  113002, 2006.

\bibitem{taminiau2008optical}
T.~H. Taminiau, F.~D. Stefani, F.~B. Segerink, and N.~F. Van~Hulst, ``Optical
  antennas direct single-molecule emission,'' \emph{Nat. Photonics}, vol.~2,
  no.~4, pp. 234--237, 2008.

\bibitem{singh2014vectorial}
A.~Singh, G.~Calbris, and N.~F. Van~Hulst, ``Vectorial nanoscale mapping of
  optical antenna fields by single molecule dipoles,'' \emph{Nano Lett.},
  vol.~14, no.~8, pp. 4715--4723, 2014.

\bibitem{farahani2007bow}
J.~N. Farahani, H.-J. Eisler, D.~W. Pohl, M.~Pavius, P.~Fl{\"u}ckiger,
  P.~Gasser, and B.~Hecht, ``Bow-tie optical antenna probes for single-emitter
  scanning near-field optical microscopy,'' \emph{Nanotechnology}, vol.~18,
  no.~12, p. 125506, 2007.

\bibitem{gross2018near}
H.~Gro{\ss}, J.~M. Hamm, T.~Tufarelli, O.~Hess, and B.~Hecht, ``Near-field
  strong coupling of single quantum dots,'' \emph{Sci. Adv.}, vol.~4, no.~3, p.
  eaar4906, 2018.

\bibitem{chugh2018flow}
D.~Chugh, J.~Wong-Leung, L.~Li, M.~Lysevych, H.~H. Tan, and C.~Jagadish, ``Flow
  modulation epitaxy of hexagonal boron nitride,'' \emph{2D Mater.}, vol.~5,
  no.~4, p. 045018, 2018.

\bibitem{dabrowska2020two}
A.~K. D{\k{a}}browska, M.~Tokarczyk, G.~Kowalski, J.~Binder, R.~Bo{\.z}ek,
  J.~Borysiuk, R.~St{\k{e}}pniewski, and A.~Wysmo{\l}ek, ``{Two stage epitaxial
  growth of wafer-size multilayer h-BN by metal-organic vapor phase epitaxy--a
  homoepitaxial approach},'' \emph{2D Mater.}, vol.~8, p. 015017, 2020.

\bibitem{tokarczyk2023effective}
M.~Tokarczyk, A.~K. D{\k{a}}browska, G.~Kowalski, R.~Bo{\.z}ek, J.~Iwa{\'n}ski,
  J.~Binder, R.~St{\k{e}}pniewski, and A.~Wysmo{\l}ek, ``{Effective substrate
  for the growth of multilayer h-BN on sapphire--substrate off-cut, pre-growth,
  and post-growth conditions in metal-organic vapor phase epitaxy},'' \emph{2D
  Mater.}, vol.~10, no.~2, p. 025010, 2023.

\bibitem{binder2023epitaxial}
J.~Binder, A.~K. Dabrowska, M.~Tokarczyk, K.~Ludwiczak, R.~Bozek, G.~Kowalski,
  R.~Stepniewski, and A.~Wysmolek, ``Epitaxial hexagonal boron nitride for
  hydrogen generation by radiolysis of interfacial water,'' \emph{Nano Lett.},
  vol.~23, no.~4, pp. 1267--1272, 2023.

\bibitem{ludwiczak2024large}
K.~Ludwiczak, A.~K. D{\k a}browska, J.~Kucharek, J.~Rogo{\.z}a, M.~Tokarczyk,
  R.~Bo{\.z}ek, M.~Gryglas-Borysiewicz, T.~Taniguchi, K.~Watanabe, J.~Binder,
  W.~Pacuski, and A.~Wysmo{\l}ek, ``{Large-area growth of high-optical-quality
  MoSe$_2$/hBN heterostructures with tunable charge carrier concentration},''
  \emph{ACS Appl. Mater. Interfaces}, vol.~16, no.~37, pp. 49\,701--49\,710,
  2024.

\bibitem{aigouy1999polarization}
L.~Aigouy, A.~Lahrech, S.~Gr{\'e}sillon, H.~Cory, A.~C. Boccara, and J.~C.
  Rivoal, ``Polarization effects in apertureless scanning near-field optical
  microscopy: an experimental study,'' \emph{Opt. Lett.}, vol.~24, no.~4, pp.
  187--189, 1999.

\bibitem{gerton2004tip}
J.~M. Gerton, L.~A. Wade, G.~A. Lessard, Z.~Ma, and S.~R. Quake, ``Tip-enhanced
  fluorescence microscopy at 10 nanometer resolution,'' \emph{Phys. Rev.
  Lett.}, vol.~93, no.~18, p. 180801, 2004.

\bibitem{preuss2022resonant}
J.~A. Preuss, D.~Groll, R.~Schmidt, T.~Hahn, P.~Machnikowski, R.~Bratschitsch,
  T.~Kuhn, S.~Michaelis~de Vasconcellos, and D.~Wigger, ``{Resonant and
  phonon-assisted ultrafast coherent control of a single hBN color center},''
  \emph{Optica}, vol.~9, no.~5, pp. 522--531, 2022.

\bibitem{dabrowska2024defects}
A.~K. D{\k{a}}browska, J.~Binder, I.~Prozheev, F.~Tuomisto, J.~Iwa{\'n}ski,
  M.~Tokarczyk, K.~P. Korona, G.~Kowalski, R.~St{\k{e}}pniewski, and
  A.~Wysmo{\l}ek, ``{Defects in layered boron nitride grown by Metal Organic
  Vapor Phase Epitaxy: luminescence and positron annihilation studies},''
  \emph{J. Lumin.}, vol. 269, p. 120486, 2024.

\bibitem{jo2023direct}
K.~Jo, E.~Marino, J.~Lynch, Z.~Jiang, N.~Gogotsi, T.~P. Darlington, M.~Soroush,
  P.~J. Schuck, N.~J. Borys, C.~B. Murray, and D.~Jariwala, ``Direct
  nano-imaging of light-matter interactions in nanoscale excitonic emitters,''
  \emph{Nat. Commun.}, vol.~14, p. 2649, 2023.

\bibitem{li2023prolonged}
S.~X. Li, T.~Ichihara, H.~Park, G.~He, D.~Kozawa, Y.~Wen, V.~B. Koman, Y.~Zeng,
  M.~Kuehne, Z.~Yuan, S.~Faucher, J.~H. Warner, and M.~S. Strano, ``Prolonged
  photostability in hexagonal boron nitride quantum emitters,'' \emph{Commun.
  Mater.}, vol.~4, p.~19, 2023.

\bibitem{noh2018stark}
G.~Noh, D.~Choi, J.-H. Kim, D.-G. Im, Y.-H. Kim, H.~Seo, and J.~Lee, ``Stark
  tuning of single-photon emitters in hexagonal boron nitride,'' \emph{Nano
  Lett.}, vol.~18, no.~8, pp. 4710--4715, 2018.

\bibitem{yu2022electrical}
M.~Yu, D.~Yim, H.~Seo, and J.~Lee, ``{Electrical charge control of h-BN single
  photon sources},'' \emph{2D Mater.}, vol.~9, no.~3, p. 035020, 2022.

\bibitem{zhigulin2023stark}
I.~Zhigulin, J.~Horder, V.~Iv{\'a}dy, S.~J.~U. White, A.~Gale, C.~Li, C.~J.
  Lobo, M.~Toth, I.~Aharonovich, and M.~Kianinia, ``Stark effect of blue
  quantum emitters in hexagonal boron nitride,'' \emph{Phys. Rev. Appl.},
  vol.~19, no.~4, p. 044011, 2023.

\bibitem{mangum2008resolving}
B.~D. Mangum, C.~Mu, and J.~M. Gerton, ``Resolving single fluorophores within
  dense ensembles: contrast limits of tip-enhanced fluorescence microscopy,''
  \emph{Opt. Express}, vol.~16, no.~9, pp. 6183--6193, 2008.

\bibitem{hoppener2009background}
C.~H{\"o}ppener, R.~Beams, and L.~Novotny, ``Background suppression in
  near-field optical imaging,'' \emph{Nano Lett.}, vol.~9, no.~2, pp. 903--908,
  2009.

\bibitem{hillenbrand2002phonon}
R.~Hillenbrand, T.~Taubner, and F.~Keilmann, ``Phonon-enhanced light--matter
  interaction at the nanometre scale,'' \emph{Nature}, vol. 418, no. 6894, pp.
  159--162, 2002.

\bibitem{huth2012nano}
F.~Huth, A.~Govyadinov, S.~Amarie, W.~Nuansing, F.~Keilmann, and
  R.~Hillenbrand, ``{Nano-FTIR absorption spectroscopy of molecular
  fingerprints at 20 nm spatial resolution},'' \emph{Nano Lett.}, vol.~12,
  no.~8, pp. 3973--3978, 2012.

\bibitem{mcleod2020multi}
A.~S. Mcleod, J.~Zhang, M.~Q. Gu, F.~Jin, G.~Zhang, K.~W. Post, X.~G. Zhao,
  A.~J. Millis, W.~B. Wu, J.~M. Rondinelli, R.~D. Averitt, and D.~N. Basov,
  ``Multi-messenger nanoprobes of hidden magnetism in a strained manganite,''
  \emph{Nat. Mater.}, vol.~19, no.~4, pp. 397--404, 2020.

\bibitem{tawfik2017first}
S.~A. Tawfik, S.~Ali, M.~Fronzi, M.~Kianinia, T.~T. Tran, C.~Stampfl,
  I.~Aharonovich, M.~Toth, and M.~J. Ford, ``{First-principles investigation of
  quantum emission from hBN defects},'' \emph{Nanoscale}, vol.~9, no.~36, pp.
  13\,575--13\,582, 2017.

\bibitem{kirchhoff2022electronic}
A.~Kirchhoff, T.~Deilmann, P.~Kr{\"u}ger, and M.~Rohlfing, ``Electronic and
  optical properties of a hexagonal boron nitride monolayer in its pristine
  form and with point defects from first principles,'' \emph{Phys. Rev. B},
  vol. 106, no.~4, p. 045118, 2022.

\bibitem{neumann2023organic}
M.~Neumann, X.~Wei, L.~Morales-Inostroza, S.~Song, S.-G. Lee, K.~Watanabe,
  T.~Taniguchi, S.~G{\"o}tzinger, and Y.~H. Lee, ``Organic molecules as origin
  of visible-range single photon emission from hexagonal boron nitride and
  mica,'' \emph{ACS Nano}, vol.~17, no.~12, pp. 11\,679--11\,691, 2023.

\end{thebibliography}


\end{document}



\author[1]{I. Niehues et al.}
\title{Nanoscale resolved mapping of the dipole emission of hBN color centers with a scattering-type scanning near-field optical microscope\\ (\textit{Supplementary Material})}
\runningtitle{Supplementary Material}

\maketitle

\vspace*{-6pt}

\section{MOVPE grown hBN}

Boron nitride was grown by Metalorganic Vapor Phase Epitaxy (MOVPE) on 2-inch sapphire c-plane substrates with a $0.6^\circ$ off-cut angle~\cite{tokarczyk2023effective}, using an {\it Aixtron CCS} 3$\times$2” system. Triethylboron and ammonia served as the boron and nitrogen precursors. A two-stage epitaxy approach~\cite{dabrowska2020two} was employed, comprising a 10-minute Continuous Flow Growth (CFG) buffer layer and a subsequent 60-minute Flow-rate Modulation Epitaxy (FME) layer growth. The initial temperature of 1300$^\circ$C (measured by an {\it ARGUS} optical pyrometer) was continuously increased until it reached 1400$^\circ$C within 2 minutes after the start of the FME stage and was maintained at this level until the end of the process. 

The samples were transferred onto other substrates by using a water-based transfer method~\cite{ludwiczak2024large}. To this end, the samples were immersed into a deionized water / isopropanol solution which releases the whole layer from the sapphire substrate. The solution can penetrate the interface between the hBN layer and the sapphire substrate, because the high-temperature growth of hBN on sapphire yields a characteristic mesh of wrinkles originating from different coefficients of thermal expansion~\cite{chugh2018flow,binder2023epitaxial}. The floating hBN layer can be transferred on other substrates by immersing the target substrate into the liquid below the floating layer and gently lifting the substrate out of the solution. The transfer process allows the hBN layer to relax and the wrinkle pattern disappears. As target substrates, commercial Si/SiO$_2$ substrates (90~nm SiO$_2$ thickness) or gold coated sapphire substrates were used. The gold layer was deposited using sputtering techniques with layer thicknesses between 45-100~nm.

\section{Scattering-type near-field optical microscope setup}
Our study employs a custom-designed scattering-type scanning near-field optical microscope ({\it neaSCOPE} from {\it neaspec/attocube}). This system is operated in tapping mode AFM using standard platinum-iridium tips that feature a 30 nm tip apex diameter ({\it Arrow-NCPt} sourced from {\it NanoWorld}), ensuring precise interaction with the sample surface. The s-SNOM setup includes a high-quality, silver-protected off-axis parabolic mirror with a numerical aperture (NA) of 0.72, which optimizes the focusing and collection efficiency of the optical system. The microscope is equipped with a 532 nm laser ({\it Cobolt 04 series}, {\it H\"ubner}), which provides the necessary excitation for photoluminescence and Raman spectroscopy. In addition, a 561~nm laser and a 633~nm laser ({\it Cobolt 08 series}, {\it H\"ubner}) are used. Integrated into the setup is an {\it Andor Kymera} spectrometer with an {\it iDus} camera, which enables detailed spectral analysis across a wide range of wavelengths.

\section{PL measurements}
The sample is initially brought into contact with the tip. The p-polarized 532 nm laser is first focused on the sample alone (without the tip involved) to maximize the Raman/PL signal, ensuring that the laser is precisely focused on the sample surface. Following this, the parabolic mirror is adjusted in the $xy$-plane to direct the laser onto the AFM tip without altering the $z$-axis positioning.

Proper alignment of the laser on the AFM tip is crucial, and this is achieved by maximizing the near-field signal (elastically scattered laser light) in the 3rd and 4th demodulation orders. It is worth noting that the near-field amplitude can be further increased by focusing the laser higher up on the tip (by moving the parabolic mirror in the $+z$-direction) than on the sample. However, it has been found that this is not the optimal position for our PL measurements.
Once the laser is correctly aligned on the AFM tip, PL spectra are recorded using the typical measurement parameters:
\begin{itemize}
	\item AFM tapping amplitude between 20~nm and 30~nm
	\item Laser power: 150~\textmu W at the tip
	\item Integration time of 0.5~seconds.
\end{itemize}
It is important that the obtained spectra include both near-field and far-field contributions and are not demodulated.

\begin{figure}[htbp]
\centering
\includegraphics[width=0.9\linewidth]{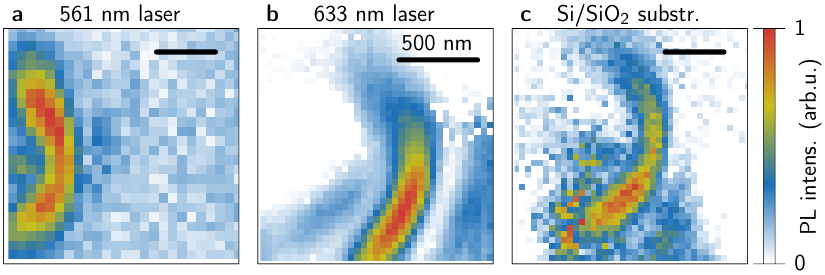}
\caption{PL maps taken with the AFM tip under varying conditions. (a, b) Excitation laser wavelength 561~nm and 633~nm, respectively, on a gold substrate (same sample as in the main text). (c)~532~nm excitation on a silicon substrate (hBN from the same growth process as the other sample but transferred to Si/SiO$_\text{2}$).}
\label{fig:S1}
\end{figure}

In Fig.~\ref{fig:S1}(a, b) PL maps with different excitation laser wavelengths are shown and (c) shows a measurement on a Si/SiO$_2$ substrate. The measurements in the main text are performed with a 532~nm excitation wavelength and on a Au substrate. 

\begin{figure}[htbp]
\centering
\includegraphics[width=0.65\linewidth]{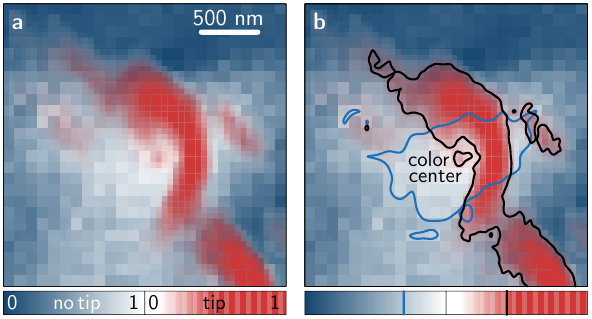}
\caption{Comparison between Figs.~1a and 1c from the main text. (a) PL image taken without AFM tip in blue overlayed with the same image taken with AFM tip in red. (b) Same as (a) but with added contour lines. The blue contour line of the far-field image is plotted at 0.7 of the global PL maximum, while the black line of the near-field image is plotted at 0.43. Note that we used a lower value to mark the dot in the near-field image due to the higher PL values in the arc. The position of the color center is marked by the isolated black solid area in the middle of the image.}
\label{fig:S2}
\end{figure}

Figure~\ref{fig:S2} shows an overlay comparison between the PL image taken without the AFM tip (blue, same as Fig.~1a in the main text) and the same image taken with the AFM tip (red, same as Fig.~1c in the main text). This again confirms the improved resolution of the spatial PL scan and the more precise localization of single color centers. The contour lines in Fig.~\ref{fig:S2}b highlight the localization of the color center in the near-field image at the isolated black solid area compared with the blue outlined area from the image without tip. Note that we used a value of 0.7 for the blue contour of the far-field image without the tip. This value can directly be referred to the overall maximum PL intensity. We used a lower value of 0.43 to mark the dot in the near-field image because the overall maximum PL value lies in the arc and the dot maximum is smaller.

\newpage

\section{Quantification of the PL enhancement}
\begin{figure}[htbp]
\centering
\includegraphics[width=0.55\linewidth]{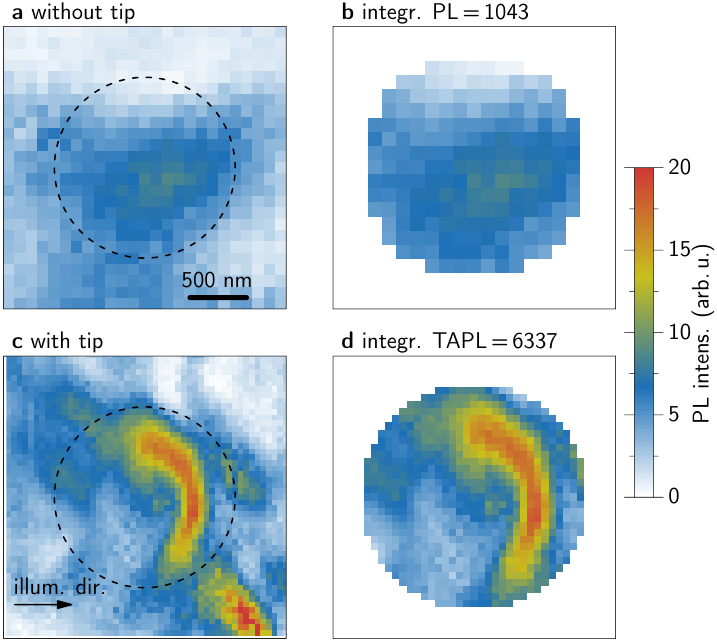}
\caption{PL intensity integration. (a, c) are the same measurements as Figs. 1a, c in the main text. (b, d) show the respective circular cuts (dashed lines in (a, c)) for spatial intensity summation.}
\label{fig:S3}
\end{figure}

In Fig.~\ref{fig:S3} we show how the spatial areas used for the spatial intensity summation are selected. (a, c) show the same results as Figs. 1a, c in the main text and the dashed circles mark the areas used for the intensity summation. (b, d) show the actual data used for summation. We need to select the same sample areas for this spatial integration because we find regions with increased count rates stemming from different emitters in the near-field image in Fig.~\ref{fig:S3}c, which we need to neglect. Finally, after summing over all pixels in (b) and (d) we find count numbers of 1043 and 6337 for PL without tip and PL with tip, respectively. This means an overall increase in detection efficiency of more than six times.

\clearpage

\section{Simulation of tip-assisted light emission}
\begin{figure}[b]
\centering
\includegraphics[width=0.65\linewidth]{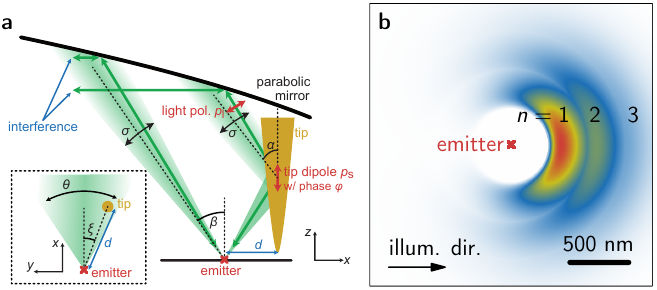}
\caption{(a) Schematic picture of the interference between direct emission from the emitter and indirect emission via scattering from the tip dipole. (b) Simulation of tip-assisted light emission with $\lambda=650$~nm, $\varphi=\pi$, $\sigma=0.1\pi$, $\beta=0.25\pi$, $\theta=0.3\pi$.}
\label{fig:S4}
\end{figure}

If we consider the situation in Fig.~\ref{fig:S4}a two beams are emitted from the color center: One directly hitting the collecting parabolic mirror, the other being scattered from the tip (experiencing a phase jump $\varphi$) and then running parallel to the first one into the mirror. These two beams will interfere constructively under the condition:
\begin{equation}
	d\left(\alpha\right)=\frac{n-\sin\left(\varphi\right)}{2\sin^3{\left(\alpha\right)}}\lambda\,,\qquad n\in\mathbb{N}\,,
\end{equation}
where $d$ is the distance between tip and emitter and $\lambda$ is the emitted photoluminescence wavelength.

The efficiency of the scattering will depend on the orientation of the light polarization $p_{\rm l}$ with respect to the symmetry axis of the tip, i.e., the orientation of the scattering dipole $p_{\rm s}$, such that the intensity of the detected light will be proportional to
\begin{equation}
	I_{\mathrm{scat}}\sim\cos^2{\left(\alpha\right)\ .}
\end{equation}
Next, we take the interference angle $\sigma$ and the focusing angle $\beta$ into account. We model this by a Gaussian intensity profile around of width $\sigma$ around $\beta$. This adds an angle-dependent intensity profile for the detection of
\begin{equation}
	I_{\mathrm{det}}^{\left(z\right)}\sim\exp{\left[-\frac{\left(\alpha-\beta\right)^2}{2\sigma^2}\right]\ .}
\end{equation}
To get a full scan of the $xy$-plane we additionally need to add the focusing as sketched in the inset. This contributes an intensity scaling with
\begin{equation}
	I_{\mathrm{det}}^{\left(xy\right)}\sim\exp{\left(-\frac{\xi^2}{2\theta^2}\right)}\ ,
\end{equation}
which roughly corresponds to the NA of the parabolic mirror.
This in-plane widening of the interference pattern additionally accounts for the non-vanishing radius of the tip.
As the mirror and tip positions are fixed and the sample is moved in the scan, the focal point moves with respect to the emitter, such that we need to add an additional intensity scaling according to this movement according to
\begin{equation}
	I_{\mathrm{focus}}^{\left(d\right)}\sim\exp{\left(-\frac{d^2}{\lambda^2}\right)}\,.
\end{equation}
For the overall intensity 
\begin{equation}	
	I_{\mathrm{PL}}\sim I_{\mathrm{scat}}I_{\mathrm{det}}^{\left(z\right)}I_{\mathrm{det}}^{\left(xy\right)}I_{\mathrm{focus}}^{\left(d\right)}
\end{equation}
we get the result in Fig.~\ref{fig:S4}b and clearly see the arc for $n=1,\ 2,\ 3$ with decreasing intensity. Note that we did not additionally include the direct emission from the color center, i.e., the central dot, in the image as it is no part of this model. The parameters are given in the caption. Arcs with higher $n$ are sometimes seen experiment if the intensity and signal-to-noise-ratio are high enough (see Fig.~2 in main text).

\section{Color center localization}
\begin{figure}[htbp]
\centering
\includegraphics[width=0.4\linewidth]{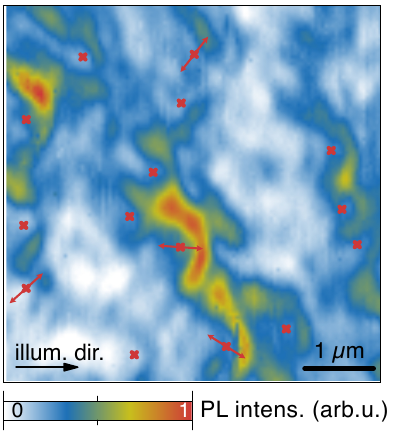}
\caption{Same PL map as in Fig.~3a in the main text with color center locations (red crosses) estimated from the arcs. The double arrows show the dipole orientation of the color centers studied in Fig.~4 in the main text.}
\label{fig:S5}
\end{figure}

Figure~\ref{fig:S5} shows the same PL map as in Fig.~3 of the main text. We include red crosses where color centers are located according to the curvatures of the arcs in TAPL. Note that the dot from TEPL is not visible for all color centers. One reason could be that the emitter is located
deep in the sample and cannot be reached with the nanofocus. In addition, we added the dipole orientations for the four color centers studied in Fig.~4 of the main text as red arrows.

\clearpage
\section{Atomic force microscopy image}
\begin{figure}[htbp]
\centering
\includegraphics[width=0.4\linewidth]{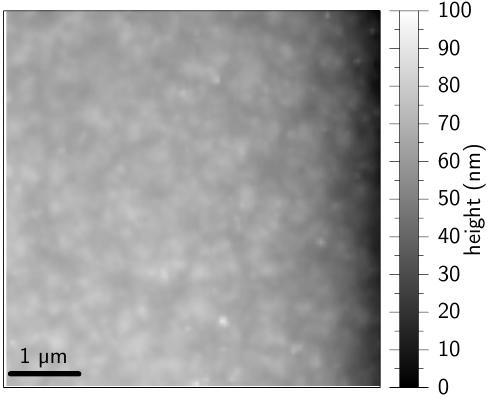}
\caption{Atomic force height profile of the sample corresponding to the PL map in Fig.~3 in the main text.}
\label{fig:S6}
\end{figure}

Figure~\ref{fig:S6} shows the height profile of the samples studied in PL in Fig.~3 of the main text. Both images were taken simultaneously. The morphology of the AFM images shows no strong inhomogeneities within the hBN flake. Therefore, from this measurement we cannot identify a correlation to the appearance of color centers and for example strain.

\section{Determining the sample rotation angle}
\begin{figure}[htbp]
\centering
\includegraphics[width=0.5\linewidth]{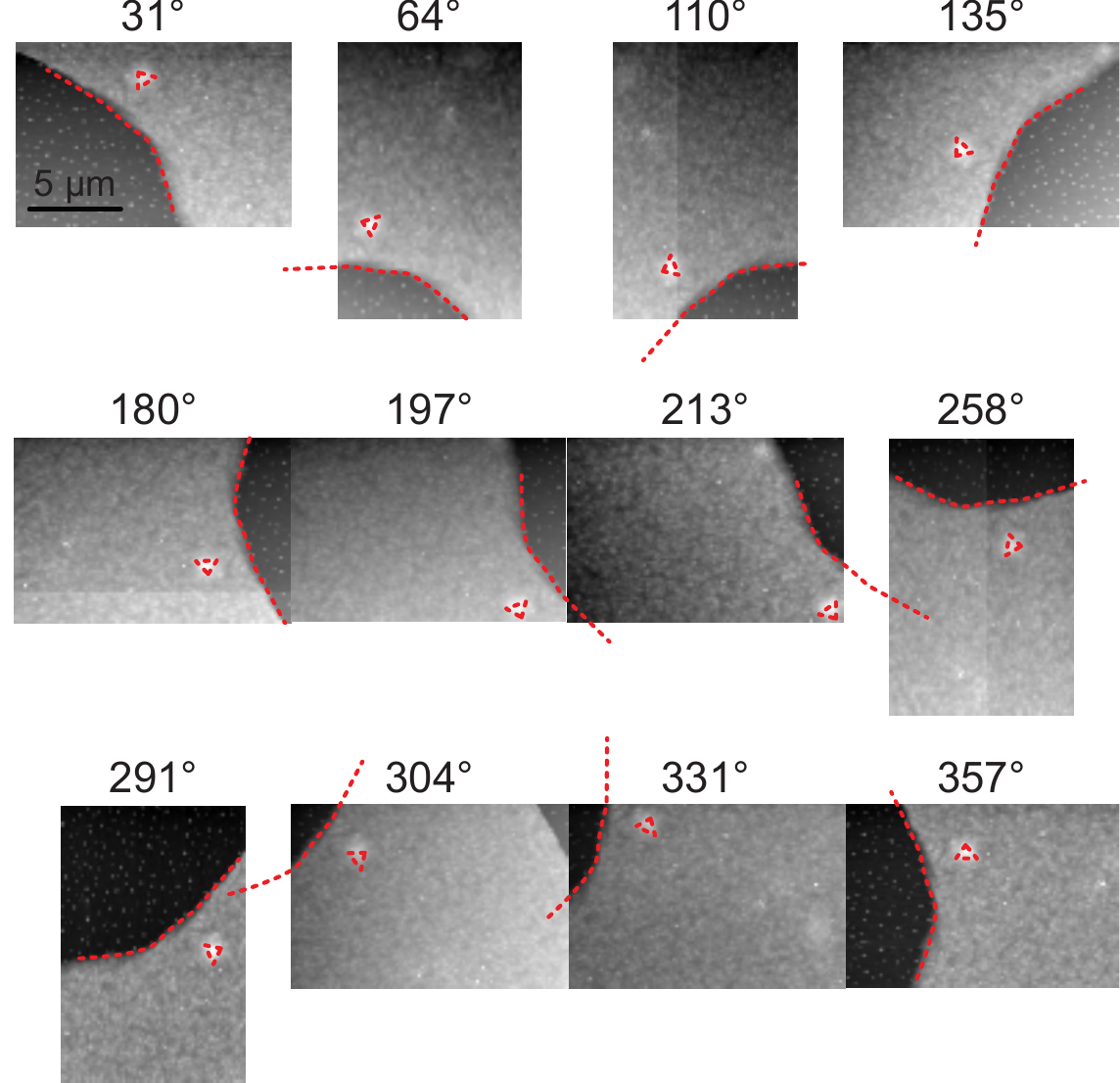}
\caption{Atomic force height profile of the sample for different sample orientations with respect to the illumination direction.}
\label{fig:S7}
\end{figure}

Figure~\ref{fig:S7} shows the height profile of the sample for all different rotation angles used in Fig.~4 of the main text. The rotation angle between sample and illumination direction was determined by tracing the flake edge (red dashed line) and a spot (most likely dirt) on the sample surface (red triangle).

